\documentclass[aps,twocolumn,superscriptaddress,floatfix,letterpaper,showpacs]{revtex4}

\usepackage{graphicx,amssymb,bm,amsfonts,natbib,amsmath,graphics,color}
\usepackage[bookmarks=false,pdfstartview=FitH,colorlinks=true,citecolor=blue,linkcolor=blue]{hyperref}

\begin{document}

\title {Widespread spin polarization effects in photoemission from topological insulators}

\author{C. Jozwiak} \email{cmjozwiak@lbl.gov} \affiliation{Advanced Light Source, Lawrence Berkeley National Laboratory, Berkeley, California 94720, USA}
\author{Y. L. Chen} \affiliation{Advanced Light Source, Lawrence Berkeley National Laboratory, Berkeley, CA 94720, USA}
\affiliation{Stanford Institute for Materials and Energy Sciences, SLAC National Accelerator Laboratory, 2575 Sand Hill Road, Menlo Park, California 94025, USA}
\affiliation{Geballe Laboratory for Advanced Materials, Departments of Physics and Applied Physics, Stanford University, Stanford, California 94305, USA}
\author{A. V. Fedorov} \affiliation{Advanced Light Source, Lawrence Berkeley National Laboratory, Berkeley, California 94720, USA}
\author{J. G. Analytis} \affiliation{Stanford Institute for Materials and Energy Sciences, SLAC National Accelerator Laboratory, 2575 Sand Hill Road, Menlo Park, California 94025, USA}
\affiliation{Geballe Laboratory for Advanced Materials, Departments of Physics and Applied Physics, Stanford University, Stanford, California 94305, USA}
\author{C. R. Rotundu} \affiliation{Materials Sciences Division, Lawrence Berkeley National Laboratory, Berkeley, California 94720, USA}
\author{A. K. Schmid} \affiliation{Materials Sciences Division, Lawrence Berkeley National Laboratory, Berkeley, California 94720, USA}
\author{J. D. Denlinger} \affiliation{Advanced Light Source, Lawrence Berkeley National Laboratory, Berkeley, California 94720, USA}
\author{Y.-D. Chuang} \affiliation{Advanced Light Source, Lawrence Berkeley National Laboratory, Berkeley, California 94720, USA}
\author{D.-H. Lee} \affiliation{Department of Physics, University of California, Berkeley, California 94720, USA} \affiliation{Materials Sciences Division, Lawrence Berkeley National Laboratory, Berkeley, California 94720, USA}
\author{I. R. Fisher} \affiliation{Stanford Institute for Materials and Energy Sciences, SLAC National Accelerator Laboratory, 2575 Sand Hill Road, Menlo Park, California 94025, USA}
\affiliation{Geballe Laboratory for Advanced Materials, Departments of Physics and Applied Physics, Stanford University, Stanford, California 94305, USA}
\author{R. J. Birgeneau} \affiliation{Materials Sciences Division, Lawrence Berkeley National Laboratory, Berkeley, California 94720, USA} \affiliation{Department of Physics, University of California, Berkeley, California 94720, USA} \affiliation{Department of Materials Science and Engineering, University of California, Berkeley, California 94720, USA}
\author{Z.-X. Shen} \affiliation{Stanford Institute for Materials and Energy Sciences, SLAC National Accelerator Laboratory, 2575 Sand Hill Road, Menlo Park, California 94025, USA}
\affiliation{Geballe Laboratory for Advanced Materials, Departments of Physics and Applied Physics, Stanford University, Stanford, California 94305, USA}
\author{Z. Hussain} \email{zhussain@lbl.gov} \affiliation{Advanced Light Source, Lawrence Berkeley National Laboratory, Berkeley, California 94720, USA}
\author{A. Lanzara} \email{alanzara@lbl.gov} \affiliation{Department of Physics, University of California, Berkeley, California 94720, USA} \affiliation{Materials Sciences Division, Lawrence Berkeley National Laboratory, Berkeley, California 94720, USA}

\date {\today}

\begin{abstract}
High resolution spin- and angle-resolved photoemission spectroscopy (spin-ARPES) was performed on the three-dimensional topological insulator Bi$_2$Se$_3$ using a recently developed high-efficiency spectrometer.
The topological surface state's helical spin structure is observed, in agreement with theoretical prediction.
Spin textures of both chiralities, at energies above and below the Dirac point, are observed, and the spin structure is found to persist at room temperature.
The measurements reveal additional unexpected spin polarization effects, which also originate from the spin-orbit interaction, but are well differentiated from topological physics by contrasting momentum and photon energy and polarization dependencies.
These observations demonstrate significant deviations of photoelectron and quasiparticle spin polarizations.
Our findings illustrate the inherent complexity of spin-resolved ARPES and demonstrate key considerations for interpreting experimental results.
\end{abstract}


\maketitle

\section{\label{sec:Intro}Introduction}

Since the first experimental observations of three-dimensional (3D) topological insulators \cite{Hsieh2008,Xia2009,Chen2009}, much attention has been turned to this new phase of condensed matter.
Generalized from the two-dimensional (2D) quantum spin Hall effect \cite{Fu2007,Moore2007}, 3D topological insulators are predicted to posses numerous novel properties including a topological magnetoelectric effect, axion electrodynamics \cite{Qi2008} and the potential for Majorana fermion physics \cite{Fu2008}.
Strong spin-orbit coupling and time-reversal symmetry are central to the topological ordering and related properties in these materials.

\begin{figure*} \includegraphics[width=17cm]{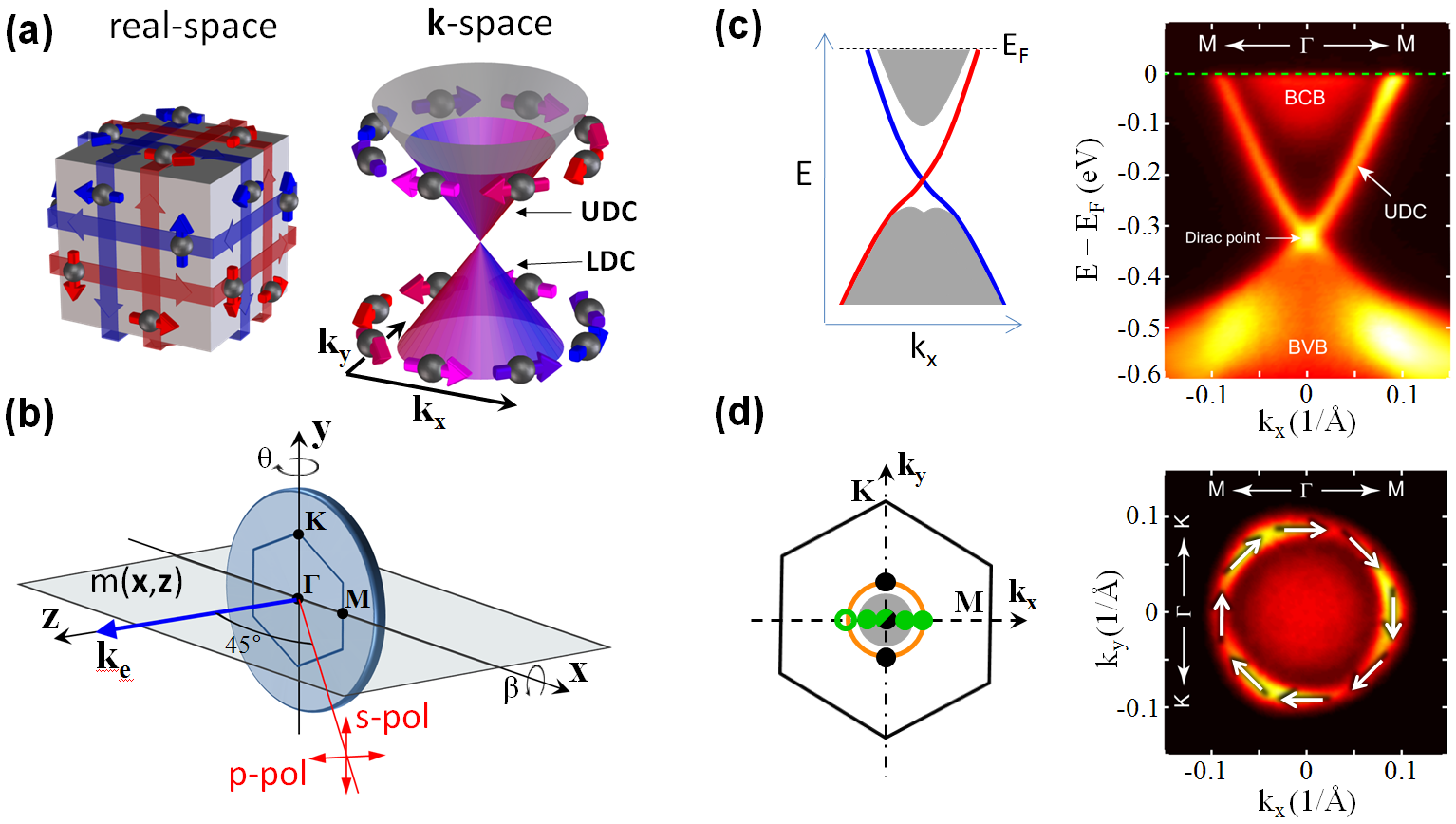}
\caption{\label{fig:diagrams}(Color) (a) Cartoon depictions of the defining spin-polarized characteristics of the topological insulator surface states, in real and momentum space.
(b) Diagram illustrating the sample and experiment geometry in the spin-resolved measurements.  Photoelectrons are collected along the vector $\mathbf{k_e}$ that is a fixed 45$^{\circ}$ from the incident photon direction, both of which are fixed in the horizontal $xz$-plane.  Spectra are taken along $\Gamma$M ($\Gamma$K) by rotating the sample about the $y$ axis ($x$ axis).
(c) ARPES intensity map of Bi$_2$Se$_3$ as a function of binding energy and momentum along the $\Gamma$M direction, taken with hv = 36 eV at 20 K.  The inset depicts the theoretical bandstructure, with the projection of the bulk bands in gray, and the TSS in blue (red) according to its spin-up (-down) polarization, quantized along the perpendicular $\mathbf{\hat{k}}_y$ direction.
(d) ARPES derived Fermi surface.  The theoretical spin polarization orientation of the TSS at $E_F$ is depicted by the white arrows.  The inset depicts the surface Brillouin zone and the TSS (orange ring) and BCB (gray circle) contributions to the Fermi surface.
}
\end{figure*}

Topological insulators are characterized by a bulk bandgap, and metallic topological surface states (TSS) of odd numbers of Dirac Fermions that can form cone-like linear dispersions in energy-momentum space \cite{Qi2010,Hasan2010}.
Such a TSS is composed of an upper Dirac cone (UDC) and lower Dirac cone (LDC) that meet at the Dirac point [Fig.~\ref{fig:diagrams}(a)].
Angle-resolved photoemission spectroscopy (ARPES), with its unique combination of energy, momentum, and surface sensitivity, is ideally suited for studying these surface electronic features, and has rapidly produced a large body of work \cite{Hsieh2008,Xia2009,Chen2009,Hsieh2009b,Chen2010,Kuroda2010a,Wray2010,Sato2010,Kuroda2010,Chen2010a,Hasan2010}.
The TSS also features unique spin-momentum locking resulting in a spin-helical texture [depicted in Fig.~\ref{fig:diagrams}(a)] that is attractive for potential application in spintronics devices.
This unusual spin texture of topological insulators makes spin-resolved ARPES (spin-ARPES) an ideal tool for studying 3D topological insulators and for identifying topological order in new materials \cite{Hsieh2009,Hsieh2009a,Nishide2010,Hirahara2010,Souma2011a,Xu2011}, despite the relative difficulty.

Due to the intrinsic inefficiency of spin-ARPES, important details of the spin texture of the TSS remain uncertain.
Measurements must be carefully made of each proposed material as these details can be material specific.
For example, the first measurements were thought to be consistent with a 100\% polarized TSS \cite{Hsieh2009a}, although the polarization of the directly measured photoelectrons did not exceed $\sim$ 20\% and were only measured along one momentum direction.
A recent first-principles calculation \cite{Yazyev2010} argues that strong spin-orbit entanglement in Bi$_2$Se$_3$ and Bi$_2$Te$_3$ greatly reduces the expected polarization of the TSS to $\sim$ 50\%, which has significant implications for device applications.
The most recent spin-ARPES measurements of photoelectron polarization, however, range from $\sim$ 60\%  \cite{Souma2011a,Xu2011} to $\sim$ 75\% \cite{Pan2011a}.
Further, the exact vectorial orientations of spin polarization in different materials, with varying out-of-plane components \cite{Fu2007,Yazyev2010,Souma2011a,Xu2011}, need further direct measurement.

Using a recently developed high-efficiency and high-resolution spin-ARPES spectrometer \cite{Jozwiak2010}, we have extensively studied the spin-polarized photoemission spectrum of the 3D topological insulator Bi$_2$Se$_3$ at both cryogenic and room temperatures, confirming key details and observing unusual spin structures not previously observed.
We demonstrate that photoemission-specific effects can lead to strong spin-polarized emission in addition to that expected from the TSS.
The results suggest that such effects must be understood and taken into account when interpreting spin-ARPES data from these exciting materials.

\section{\label{sec:Exp}Experimental Details}

The experiments were performed on Bi$_2$Se$_3$ crystals grown as discussed elsewhere \cite{Chen2009,Chen2010} and cleaved \textit{in-situ} along the (111) plane at $\sim$20 K in vacuum of $5\times 10^{-11}$ torr.
All data were taken at $\sim$20 K, except for Fig.~\ref{fig:SDb}(c) which was measured at room temperature. 
The high-resolution, spin-integrated ARPES data in Figs.~\ref{fig:diagrams}(c)~and~\ref{fig:diagrams}(d)  were taken at beamline 10.0.1 of the Advanced Light Source, Berkeley.
The data in Figs.~\ref{fig:SDa}--\ref{fig:cores} and \ref{fig:hv70} were taken at beamline 12.0.1.1 using linear, $p$-polarized light, while the data in Fig.~\ref{fig:merlin} were taken at beamline 4.0.3 using linear $p$- and $s$-polarized light.
The spin-resolved photoemission experiments were performed with an in-house developed high-efficiency spectrometer based on low-energy exchange scattering and time-of-flight (TOF) techniques.
The instrument and data acquisition procedures are discussed in detail in Refs.~\onlinecite{Jozwiak2010,Jozwiakthesis}.
The total combined experiment energy and angle resolutions were $<$~30~meV and $\pm$1$^{\circ}$, respectively.
The experiment geometry is illustrated in Fig.~\ref{fig:diagrams}(b).

\section{\label{sec:Results}Results and Discussion}

Figs.~\ref{fig:diagrams}(c)~and~\ref{fig:diagrams}(d) show high-resolution, spin-integrated ARPES data taken from the Bi$_2$Se$_3$ sample with a traditional hemispherical analyzer.
Panel (c) shows the energy-momentum dispersions measured along $\Gamma$M next to a corresponding cartoon depiction of the predicted bandstructure.
Qualitative agreement is found, with the BCB, BVB, and crossing TSS bands clearly observed.
The fact that a significant portion of the BCB is occupied and observed in the ARPES data signifies that the sample is n-doped.
The UDC is clearly observed as a sharp and distinct dispersion in the data, and hence is expected to be spin polarized accordingly.
Although hints of an LDC are also visible, it appears outside the bulk bandgap and within the BVB; hybridization is likely, and whether the hybridized LDC maintains the predicted TSS helical spin polarization is unclear.
Panel (d) shows the measured Fermi surface consisting of a nearly circular ring (due to the UDC) and a central mass of spectral weight (due to the BCB).
These observations are in agreement with previous measurements \cite{Xia2009,Chen2010}.

\begin{figure} \includegraphics[width=8.5cm]{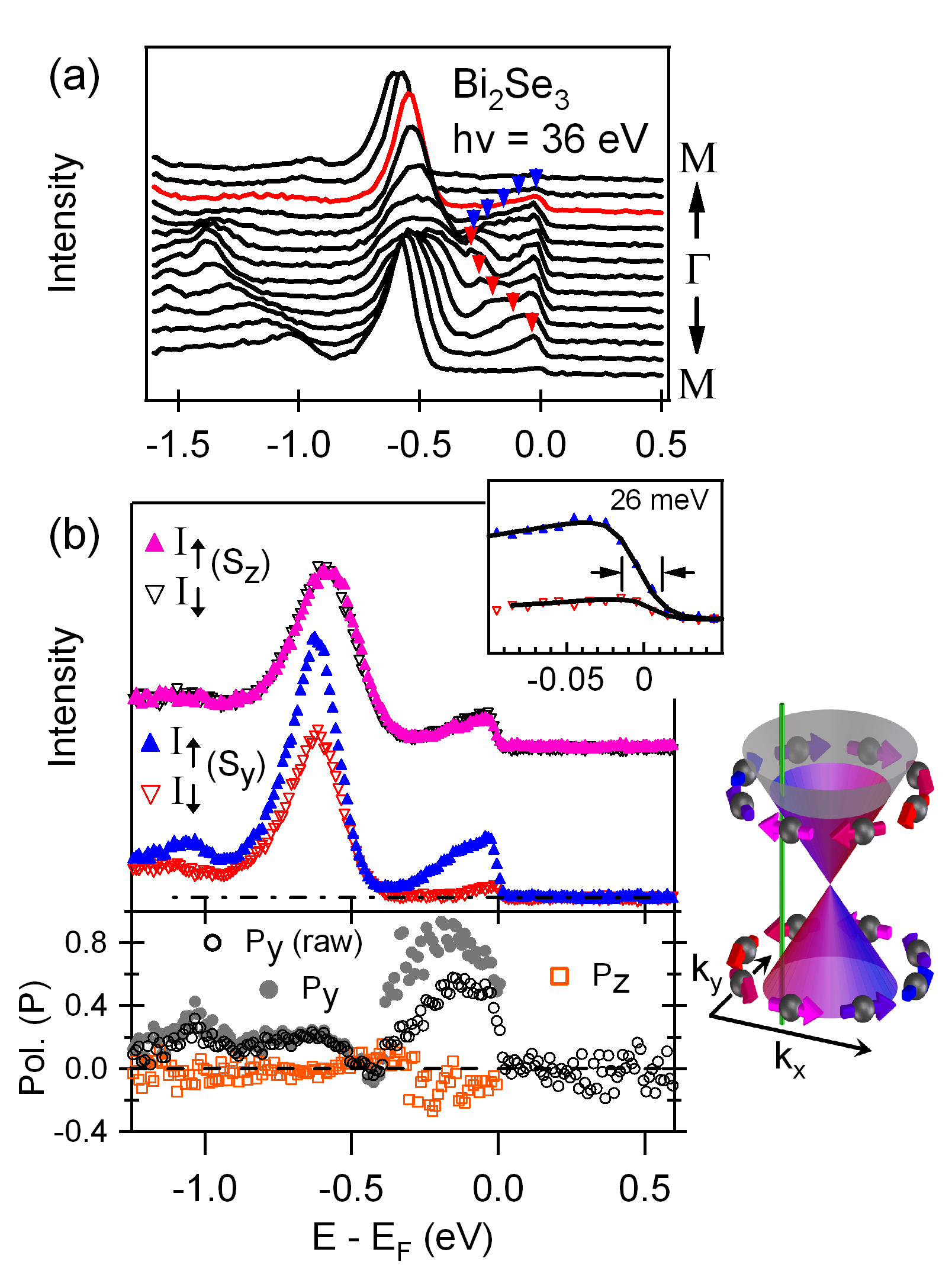}
\caption{\label{fig:SDa}(Color) (a) Spin-integrated EDCs as a function of momentum along $k_x$ ($\Gamma$M) of Bi$_2$Se$_3$.  UDC peaks are marked by blue (spin-up) and red (spin-down) arrows according to the predicted $P_y$.  Other peaks are related to bulk bands.
(b) Spin-resolved EDCs corresponding to the red EDC in (a), with the spin quantization axis along the out-of-plane $\mathbf{\hat{z}}$ direction (upper), and the in-plane $\mathbf{\hat{y}}$ direction (lower).  The EDC location is depicted by the vertical green line in the TSS cartoon inset, and the green circle in the Fig.~\ref{fig:diagrams}(d) inset.  The corresponding spin polarization curves are shown in the bottom panel.  Open black circles are the raw $P_y$, while the solid gray circles correspond to the effective polarization once a constant background (dash-dot line in upper panel) is removed.  Inset shows close-up of spin-resolved Fermi level with Fermi edge fits.
}
\end{figure}

Figure~\ref{fig:SDa} shows data taken with the spin-resolved spectrometer.
Panel (a) shows a stack of spin-integrated energy distribution curves  (EDCs) taken along $\Gamma$M (the $k_x$ axis as defined in the present geometry) in the rapid-acquisition mode of the spectrometer \cite{Jozwiak2010}.
Several features are visible near the Fermi level, including the surface state and bulk valence and conduction band peaks.
The triangles mark the dispersing peaks of the TSS UDC with the blue (red) color corresponding to the predicted spin polarization directed along the $y$-axis.
These dispersing peaks meet at $\Gamma$ at $\sim$ 300 meV binding energy, forming the Dirac point.
The much larger peaks just beyond the Dirac point are largely due to the BVB, as they depend strongly on incident photon energy and their binding energies are outside the known bulk bandgap \cite{Xia2009,Chen2010}.
The relative high intensity of the BVB peaks masks possible features related to the LDC.
Finally, the peaks near $E_F$ at $\Gamma$ within the UDC originate from the occupied states of the BCB.

Panel (b) shows the high-resolution spin-resolved EDC corresponding to the red EDC in panel (a), away from $\Gamma$ along $k_x$ and near $k_F$ of the UDC.
The EDC is resolved into separate intensity channels of spin-up (I$_\uparrow$) and spin-down (I$_\downarrow$) photoelectrons.
The upper pair (pink/black) uses a spin quantization axis along the out-of-plane $\mathbf{\hat{z}}$ direction, while the lower pair (blue/red) uses a quantization axis along the vertical in-plane $\mathbf{\hat{y}}$ direction, perpendicular to the in-plane momentum direction, $\mathbf{\hat{k}}_x$.
The inset shows a close-up view of the Fermi level of the spin-resolved EDCs, indicating the high energy resolution achieved, extracted as the width of the spin-resolved Fermi edge.

It is immediately clear from these EDCs that there is very little dependence on $S_z$, but significant dependence on $S_y$, where $S_z$ ($S_y$) is the $\mathbf{\hat{z}}$ ($\mathbf{\hat{y}}$) component of photoelectron spin.
As these EDCs are taken in the vicinity of the UDC $k_F$, the intensity at $E_F$ is primarily from the UDC, and its strong +$S_y$ character agrees well with the predicted spin-texture of the TSS (see inset).
However, the large $S_y$ dependence of the strong peak near $E=-0.6$ eV in the BVB is surprising; we are not aware of other measurements or predictions for such polarization of BVB quasiparticles in these materials.
If the measured spin dependence were due to the LDC of the TSS, it should instead be polarized opposite to the UDC (see inset).
As we see below, this striking behavior is due to the possible inequivalence of the spin polarization of the quasiparticles within a material and that of the free photoelectrons that are measured \cite{Henk2003,Kuemmeth2009}.
Obviously, such behavior should be understood before spin-resolved ARPES data can be correctly interpreted.

The lower panel of Fig.~\ref{fig:SDa}(b) plots the corresponding photoelectron spin polarizations, $P_{z,y}=\frac{I_\uparrow - I_\downarrow}{I_\uparrow +I_\downarrow}$.
The open black circles mark the polarization, $P_y$ along the $y$ axis, as directly extracted from the measurement as is customary in spin polarimetry (e.g. Ref.~\onlinecite{Jozwiak2010} and references therein).
It shows a clear peak near $E_F$, again in line with the expected polarization of the UDC, and a smooth positive feature through the BVB at $E \lesssim -0.5$~eV.

\begin{figure*}\includegraphics[width=18cm]{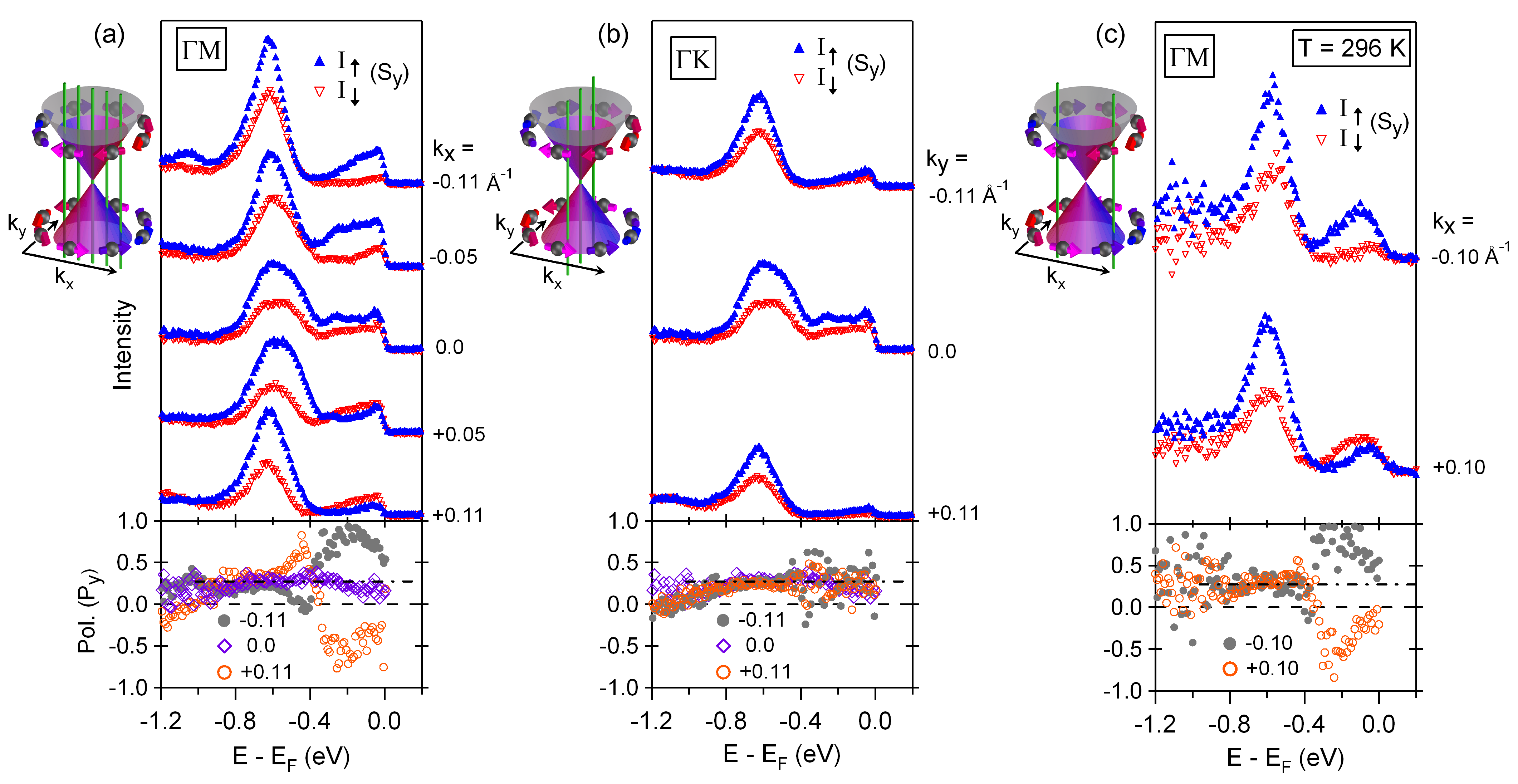}
\caption{\label{fig:SDb}(Color) (a) Spin-resolved EDCs along $k_x$ ($\Gamma$M).  Approximate locations are shown by vertical green lines in left insets and corresponding green circle and dots in the Fig.~\ref{fig:diagrams}(d) inset.  Spin polarization curves for the two extreme $k$-space locations at $\pm k_x$ and at $\Gamma$ are shown in the lower panel.
(b) Same as for (a), except along $k_y$ ($\Gamma$K).
(c) Same as for (a), except at room temperature.  All other panels are at 20 K.}
\end{figure*}

The flat, nonzero intensity signal in the EDCs above $E_F$, and the sharp drop in $P_y$ to a constant $P_y=0$ above $E_F$, show that there is a constant and unpolarized background signal emitted by the higher-order light passed by the beamline monochromator.
This unpolarized background can mask the polarization of photoelectrons emitted by first-order light by decreasing the polarization of the total photocurrent at all energies, and hence should be removed before the polarization is extracted. 
There may be additional inelastic background contributions at binding energies below $E_F$, but the high precision of the above $E_F$ measurement allows the accurate subtraction of this minimum background which is assured to be unpolarized.
The solid gray dots show $P_y$ after this background is subtracted.
Note that this procedure has no impact on the spin-resolved EDCs (upper panel).
It also has little impact on the polarization curves at energies where the signal intensity is high relative to the background (e.g., the BVB), but results in significant increase at energies of lower relative intensity (e.g., the UDC).
Finally, due to the subtraction leaving ${I_\uparrow +I_\downarrow} =0$ above $E_F$, the polarization is undefined above $E_F$, and so these data points are removed.
Each polarization curve in the rest of the present work follows this background removal.

The background corrected $P_y$ curve shows a high degree of spin polarization associated with the TSS of $\gtrsim$ 80\%.
This is higher than previous spin-resolved ARPES measurements of TSSs \cite{Hsieh2009a,Nishide2010,Hirahara2010,Souma2011a,Xu2011,Pan2011a}.
 This may result from a combination of improved experimental resolution and the described background subtraction, which is enabled by the precise measurement of the above $E_F$ signal.
It is also higher than the $\sim$ 50\% recently predicted by first principles calculations \cite{Yazyev2010}, which may suggest that understanding is not yet complete.
In contrast to $P_y$, the $P_z$ curve shows no polarization in the BVB, and a small, but non-zero out-of-plane polarization component near $E_F$.

The present measurement of $P_y \gtrsim 80\%$ in the UDC and the unexpected $P_y\gtrsim 20\%$ in the BVB are not due to an instrumental, or `false', asymmetry associated with the spectrometer, because this was previously measured to be less than 0.035\% \cite{Jozwiak2010}.
This is also seen in the present data in Fig.~\ref{fig:SDa}(b): First, the above $E_F$ signal is completely unpolarized as expected for a nonmagnetic material; and second, any instrumental asymmetry in the measurement of $P_y$ would be similarly present in the measurement of $P_z$ due to details of the spectrometer design \cite{Jozwiak2010}, and $P_z$ measures extremely close to zero in the BVB.

Figure~\ref{fig:SDb} presents similar EDCs at various momenta, with spin polarization analyzed along $\mathbf{\hat{y}}$.
Panel (a) shows spin-resolved EDCs at several values of $k_x$ along $\Gamma$M.
Strongly contrasting spin-dependent behaviors between the TSS and bulk band features are observed as a function of crystal momentum.
The strong $+S_y$ character near $E_F$ for $k_x<0$ flips to primarily $-S_y$ character for $k_x>0$.
This agrees with the helical spin structure of the UDC of the TSS (see inset).
In fact, time-reversal symmetry (TRS) in these materials requires $P_{\textrm{qp}}(\mathbf{-k})=-P_{\textrm{qp}}(\mathbf{k})$, where $P_{\textrm{qp}}(\mathbf{k})$ is the spin polarization of a quasiparticle at momentum $\mathbf{k}$.
This is in stark contrast to the BVB feature, which has nearly constant $+S_y$ character independent of $k_x$.
This appears to give the entire EDC at $k_x=0$ a nonzero spin dependence despite TRS which requires quasiparticles at $\mathbf{k}=0$ to have $P=0$.

These characteristics and more are visible in the corresponding $P_y$ curves in the bottom panel.
The $P_y(k_x=0)$ curve (purple diamonds) is smooth with little variation from $\sim$+25\% (marked by the horizontal dash-dot line), roughly centered near the BVB intensity peak.
The $P_y(k_x=\pm0.11$ \AA$^{-1}$) curves (orange circles and gray dots) qualitatively appear reflected about the polarization at $k_x=0$.
Indeed, they are nearly identical in the BVB range, but have large and opposing peaks at $E \sim -0.2$ eV due to the intrinsic spin texture of UDC quasiparticles.
These two curves also have smaller and oppositely directed peaks at $E \sim -0.45$ eV.
The opposite signs (referenced to the dash-dot line) of this pair of peaks compared to the pair due to the UDC, is a strong sign that they are signatures of the LDC and its predicted opposite spin texture chirality (see inset).
These peaks may also be due to possibly strong hybridization between the TSS and bulk bands \cite{Checkelsky2009}.
So although the EDCs do not contain obvious evidence that the spin helical texture of the TSS persists below the Dirac point and out of the bulk bandgap in Bi$_2$Se$_3$, these $P_y$ curves offer the first observation that it does.
Similar observations were recently made in BiTlSe$_2$ \cite{Xu2011a}.
It is remarkable how clearly the two contrasting spin behaviors---the polarization component that is independent of $k_x$, and the intrinsic TSS spin features that are strongly dependent on $k_x$---are displayed in these $P_y$ curves.
 
Panel (b) presents a similar stack of spin-resolved EDCs, but at several values of $k_y$ along $\Gamma$K.
Here there is no reversal of $S_y$ character at $\pm k_y$ in the energy range of the TSS, again in agreement with the helical spin structure of the TSS that has $P_y=0$ at $k_x=0$, independent of $k_y$ [see inset and Fig.~1(d)].
The $P_y(k_y$=$\pm$0.11 \AA$^{-1}$) curves closely follow the $P_y$($\mathbf{k}$=0) curve of roughly +25\% through the entire energy range.
Note that although there is a relative increase in noise in the $P_y (k_y=\pm 0.11$ \AA$^{-1})$ curves at $E>-0.4$ eV due to the relative decrease in photoemission intensity, the average value plainly follows the $P_y(\mathbf{k}$=0) curve.
Clearly, the component of $P_y$ that is independent of $k_x$ in panel (a), is also independent of $k_y$.

One of the most exciting aspects of 3D topological insulators is the possibility for topological effects, which were once thought to require extreme cryogenic temperature, to exist even at room temperature \cite{Moore2009,Moore2010}.
Similar to Fig.~\ref{fig:SDb}(a), panel (c) shows spin-resolved EDCs at $\pm$$k_x$ along $\Gamma$M, but with the sample at room temperature rather than the cryogenic temperatures of the rest of the paper ($\sim$ 295 K vs 20 K).
Both the $k_x$-dependent and -independent spin polarization features observed in panel (a) are still present.
The $k_x$ independent feature of $P_y$ in the BVB remains $\sim25\%$, and so appears temperature independent.
The large $|P_y|$ of the UDC also persists.
This is the first direct measurement of the TSS spin-helical texture at room temperature, paving the way for room-temperature spintronics applications.

In summary, Figs.~\ref{fig:SDa}~--~\ref{fig:SDb} uncover two strikingly different spin effects in terms of momentum dependence.
The strongly $\mathbf{k}$-dependent polarization features are understood in terms of the intrinsic quasiparticle spin structure of the TSS.
The $\mathbf{k}$-independent features cannot be similarly understood; such a quasiparticle spin structure, with $P(\mathbf{k})=P(-\mathbf{k})$, breaks TRS and results in a net imbalance of spin, which is unlikely as the material is not magnetic.
Since this polarization cannot be explained by instrumental asymmetries as discussed above, it must be due to significant inequivalence between quasiparticle and photoelectron spin.

The well-known Fano effect \cite{Fano1969} is a classic example of obtaining spin polarized photoelectrons from unpolarized atomic initial states using circularly polarized light.
Less intuitively, photoelectrons emitted from unpolarized atomic subshells of orbital angular momentum $l>0$ into well-defined angular directions can be spin polarized even when using linear and unpolarized light \cite{Lee1974,Cherepkov1979,Heinzmann1979}.
This general effect of the photoconversion process inducing a difference in spin polarization between the initial state and the photoelectron is not due to the photon operator altering electron spins; in the dipole approximation it does not act on spin.
Instead it is the result of spin-dependent photoemission dipole matrix elements (SMEs).

Although SMEs can occur with circular and unpolarized light, we focus here on linearly polarized light.
In this case, the SME-induced photoelectron polarization vector for photoemission from atoms is given by \cite{Lee1974,Cherepkov1979}
\begin{equation}\label{eqn:psme}
\boldsymbol{\vec P}_{\mathrm{SME}} = \frac{2\xi \left( \boldsymbol{\hat{k}}_e \cdot \boldsymbol{\hat{\epsilon}} \right) }{1+\beta\left(\frac{3}{2}(\boldsymbol{\hat{k}}_e \cdot \boldsymbol{\hat{\epsilon}})^2 - \frac{1}{2}\right)}\left[ \boldsymbol{\hat{k}}_e \times \boldsymbol{\hat{\epsilon}} \right] \,,
\end{equation}
where $\boldsymbol{\hat{k}}_e$ and $\boldsymbol{\hat{\epsilon}}$ are the outgoing photoelectron and photon polarization unit vectors, respectively.
The denominator of Eq.~(\ref{eqn:psme}) is the usual expression for the angular distribution of photoemission where $\beta$ is the asymmetry parameter.
The parameter $\xi$ reflects the interference between the possible $l+1$ and $l-1$ continuum photoelectron states, which is the source of the spin dependence.
Thus, this SME-induced polarization is due to the spin-orbit interaction, and is dependent on details of the initial and final photoelectron states, and therefore also photon energy.
Equation~(\ref{eqn:psme}) also shows that the magnitude and orientation of $\boldsymbol{\vec P}_\mathrm{SME}$ are dependent on and determined by, respectively, the relative orientation of the photon polarization and the outgoing photoelectrons, as dictated by symmetry.
Similar to the spin-orbit induced spin polarization phenomena in electron scattering \cite{Kesslerbook}, parity conservation requires $\boldsymbol{\vec P}_\mathrm{SME}$ to be perpendicular to the reaction plane formed by $\boldsymbol{\hat{\epsilon}}$ and $\boldsymbol{\hat{k}}_e$, or more generally to any mirror planes of the complete system.
In the specific cases of $s$-polarized light where $\boldsymbol{\hat{k}}_e \cdot \boldsymbol{\hat{\epsilon}} =0$, and $p$-polarized light with $\boldsymbol{\hat{k}}_e \times \boldsymbol{\hat{\epsilon}} =0$, there are two orthogonal mirror planes, and therefore $\boldsymbol{\vec P}_\mathrm{SME}$=0.

\begin{figure} \includegraphics[width=8.8cm]{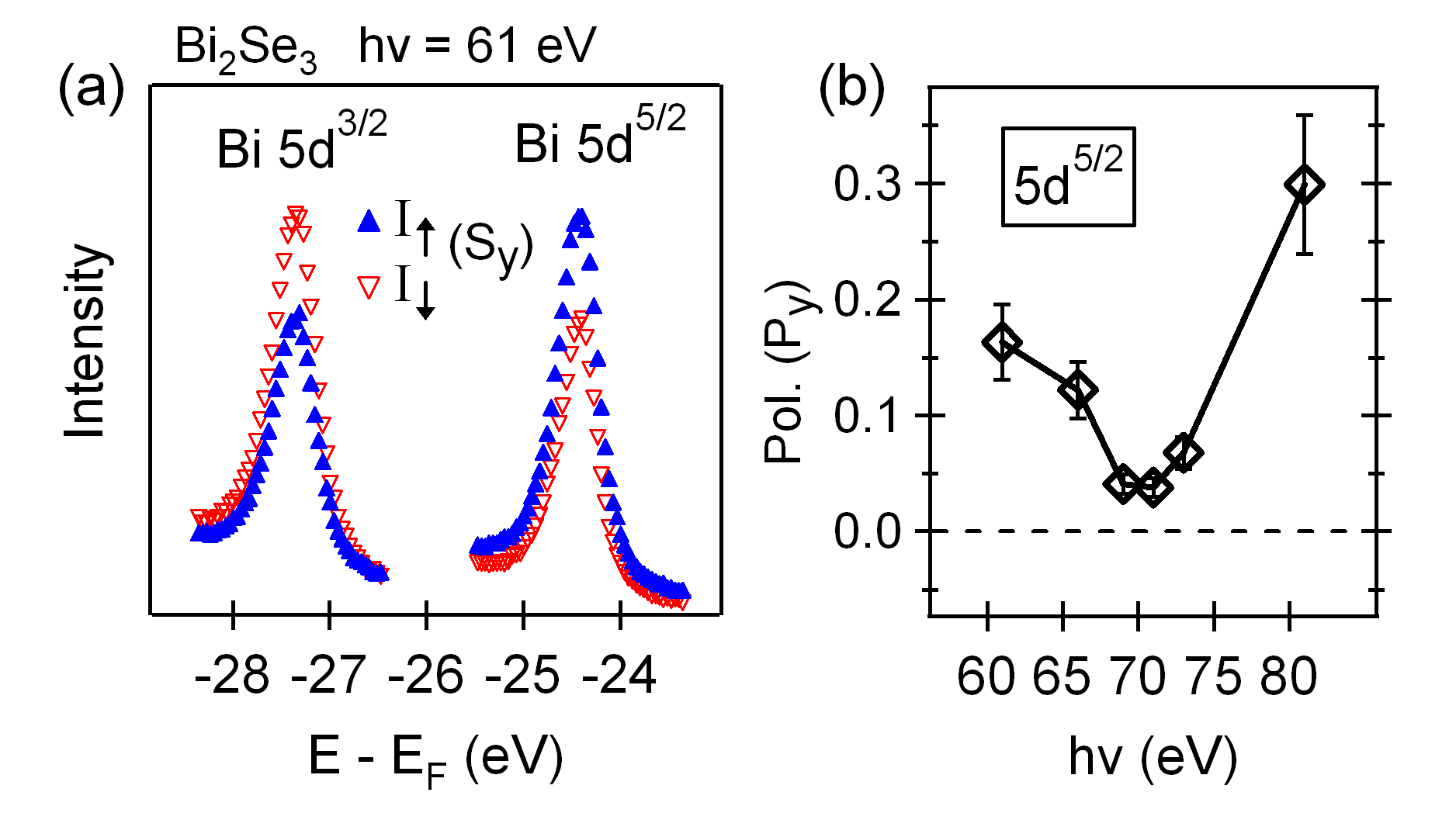}
\caption{\label{fig:cores}(Color online). (a) Spin-resolved EDCs of the Bi 5d core levels.
(b) Polarization (integrated across width of peak) of the 5d$^{5/2}$ core level, as a function of incident photon energy.
}
\end{figure}

This general effect also occurs in the solid state, clearly observed here in Bi$_2$Se$_3$ in the spin polarized photoemission from the shallow Bi $5d$ core levels.
Figure~\ref{fig:cores}(a) shows spin-resolved spectra of the $5d^{3/2}$ and $5d^{5/2}$ levels, taken at normal emission with the spin quantization axis along the $\mathbf{\hat{y}}$ direction, perpendicular to the reaction plane.
Although the initial states are not spin polarized, a large $P_y$ is observed for both peaks.
The spin-orbit origin of the polarization is evident by the opposing polarization of the two fine structure levels.
This polarization also has a strong photon energy dependence, shown in Fig.~\ref{fig:cores}(b), a characteristic common to photoemission matrix element effects.

Such SME-induced polarization has also been observed in photoemission from other nonmagnetic solid state core levels, including the Cu $2p$ and $3p$ (Ref.~\onlinecite{Roth1994}), W $4f$ (Ref.~\onlinecite{Rose1996}), and Pt $4d$ and $4f$ levels (Ref.~\onlinecite{Yu2008}).
In agreement with Eq.~(\ref{eqn:psme}) and the above symmetry considerations, the polarization vectors are perpendicular to the reaction plane when the photons are $p$-polarized \cite{Roth1994,Rose1996,Yu2008}, and the polarization is zero when the photons are $s$-polarized ($\boldsymbol{\hat{k}}_e \cdot \boldsymbol{\hat{\epsilon}} =0 $) \cite{Rose1996}.
While some quantitative differences are observed in the solid state \cite{Roth1994,Rose1996,Roth1997}, these polarization phenomena are clearly related to the atomic description and follow the same form as Eq.~(\ref{eqn:psme}).
To our knowledge, the Bi $5d$ levels of the present measurement are the shallowest core levels reported to display such SME-induced polarized emission.

SMEs in the solid state are not confined to core levels, and have been predicted \cite{Tamura1987,Tamura1991,Tamura1991a,Henk1994} and subsequently observed \cite{Schmiedeskamp1988,Schmiedeskamp1991,Irmer1992,Irmer1994,Irmer1995,Irmer1996} in various related forms in the valence bands of Pt and Au single crystals.
As seen in various Rashba-Bychkov spin-split surface states \cite{Hochstrasser2002,Hoesch2004,Shikin2008}, TRS in these systems requires that any intrinsic spin polarization of quasiparticles, $\boldsymbol{\vec P}_{\textrm{qp}}$, must be antisymmetric with respect to $\boldsymbol{\vec k}_{\parallel}=0$, where $\boldsymbol{\vec k}_{\parallel}$ is the quasiparticle in-plane crystal momentum.
This ensures zero net spin imbalance integrated over all $k$-space and that $\boldsymbol{\vec P}_{qp}(\boldsymbol{\vec k}_{\parallel}=0)=0$.
In each of these cases \cite{Schmiedeskamp1988,Schmiedeskamp1991,Irmer1992,Irmer1994,Irmer1995,Irmer1996}, however, normal emission ($\boldsymbol{\vec k}_{\parallel}=0$) photoelectrons from valence bands were found to be significantly spin polarized ($\sim$ 10 - 20\%) dependent on the photon energy and polarization.
These effects were explained in terms of SMEs.
It should be noted that no such induced polarizations are predicted within the `three-step model' of photoemission \cite{Borstel1982} due to the inversion symmetry of the assumed infinite crystal; they are only contained within more realistic `one-step model' calculations where initial and final states are defined within a semi-infinite half-space \cite{Tamura1987,Tamura1991,Tamura1991a,Schmiedeskamp1988,Irmer1992}.

In the present work, the $\mathbf{k}$-independent polarization features in Figs.~\ref{fig:SDa}~and~\ref{fig:SDb} can be similarly understood in terms of SME effects as spin-orbit coupling is very strong in the topological insulators.
The fact that this polarization effect has a large $\mathbf{\hat{y}}$ component and zero $\mathbf{\hat{z}}$ component [see the BVB peak in Fig.~\ref{fig:SDa}(b)] agrees with the form of Eq.~(\ref{eqn:psme}) requiring that any $\boldsymbol{\vec P}_{\textrm{SME}}$ must be perpendicular to the reaction plane (here the $xz$-plane, see Fig.~\ref{fig:diagrams}).
In Fig.~\ref{fig:SDb}, $k$-space is scanned by rotating the sample, while leaving the orientation between the incident light, $\boldsymbol{\hat{\epsilon}}$, and the photoelectron collection direction, $\boldsymbol{\vec k_e}$, fixed, thus keeping $\boldsymbol{\vec P}_{\textrm{SME}}$ nearly constant.
This is consistent with the observed $\mathbf{k}$-independence of this polarization feature.
As in Refs.~\onlinecite{Tamura1987,Tamura1991,Tamura1991a,Henk1994}, fully relativistic one-step photoemission calculations are required for quantitative prediction and comparison.

 \begin{figure} \includegraphics[width=6cm]{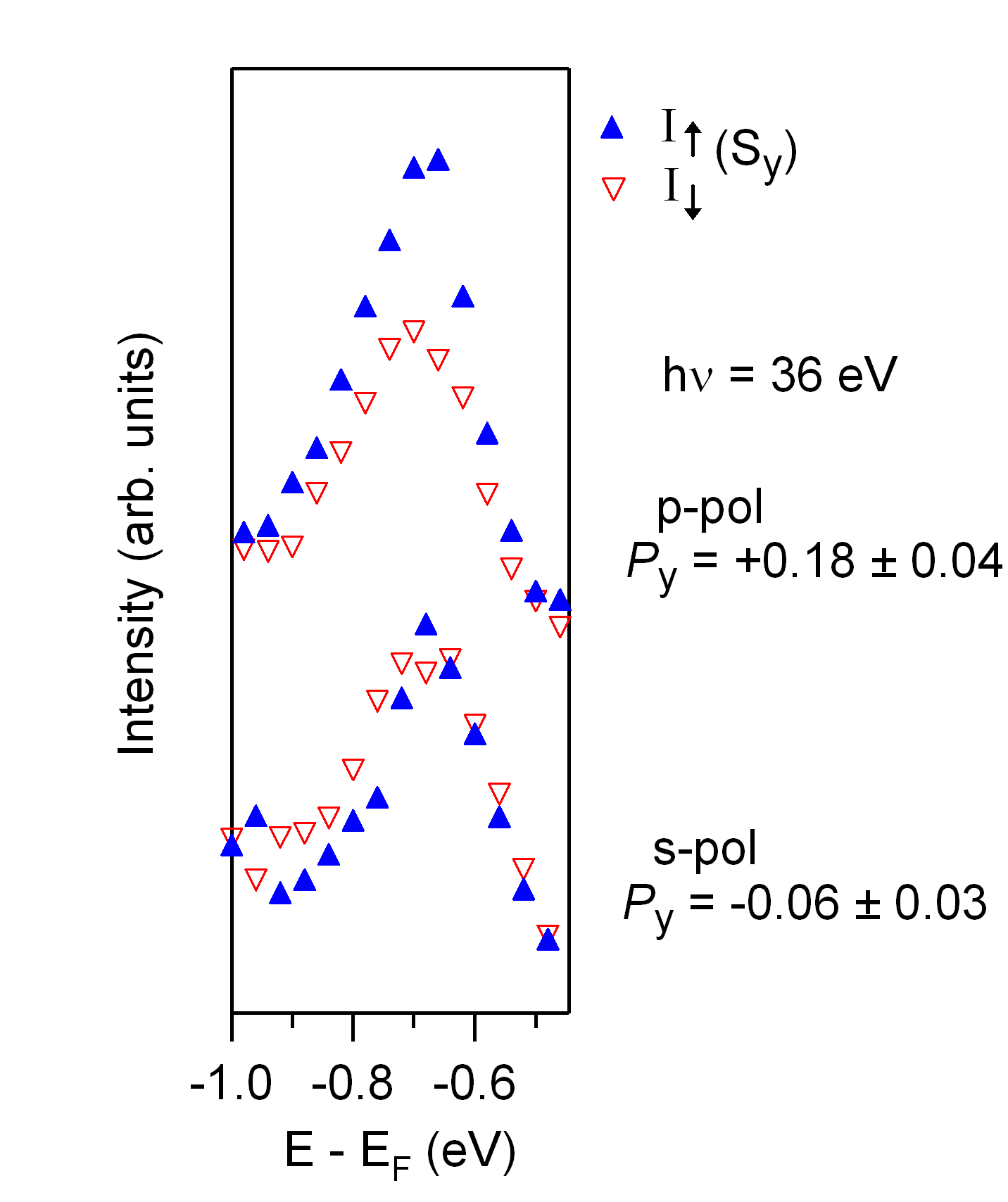}
\caption{\label{fig:merlin}(Color online). Spin-resolved EDCs of VB peak, taken along $k_x$ ($\Gamma$M), similar to Fig.~\ref{fig:SDa}(b).  Spin is measured along $\mathbf{\hat{y}}$.  The upper (lower) EDC is taken with $p$-polarized ($s$-polarized) light with $\boldsymbol{\hat{\epsilon}}$ within the $xz$-plane ($yz$-plane).  The reported $P_y$ values for each are integrated through the full energy range shown.
}
\end{figure}

Further support of an SME origin of the $\mathbf{k}$-independent polarization feature can be found in its dependence on the incident photon polarization.
Figure~\ref{fig:merlin} compares spin-resolved EDCs taken with $p$- and $s$-polarized light (see Fig.~\ref{fig:diagrams}).
The $k$-space location matches that of Fig.~\ref{fig:SDa}(b), and the energy range is now centered on the BVB peak which exhibits the $\mathbf{k}$-independent spin polarization.
Again, when excited with $p$-polarized light, as in Figs.~\ref{fig:SDa}~and~\ref{fig:SDb}, the BVB photoelectrons are observed with a significant $P_y$.
In contrast, when excited with $s$-polarized light ($\boldsymbol{\hat{k}}_e \cdot \boldsymbol{\hat{\epsilon}} =0 $), no clear $P_y$ is measured.
In the latter case, with $\boldsymbol{\hat{\epsilon}}$ completely in the surface plane, the photoemission cross section of the BVB states, which have mainly out-of-plane directed $p_z$-orbital character \cite{Zhang2009}, is reduced by over an order of magnitude, explaining the reduced statistics compared with the previous figures.
Nevertheless, a large change in $P_y$ is clear.

 \begin{figure*} \includegraphics[width=18cm]{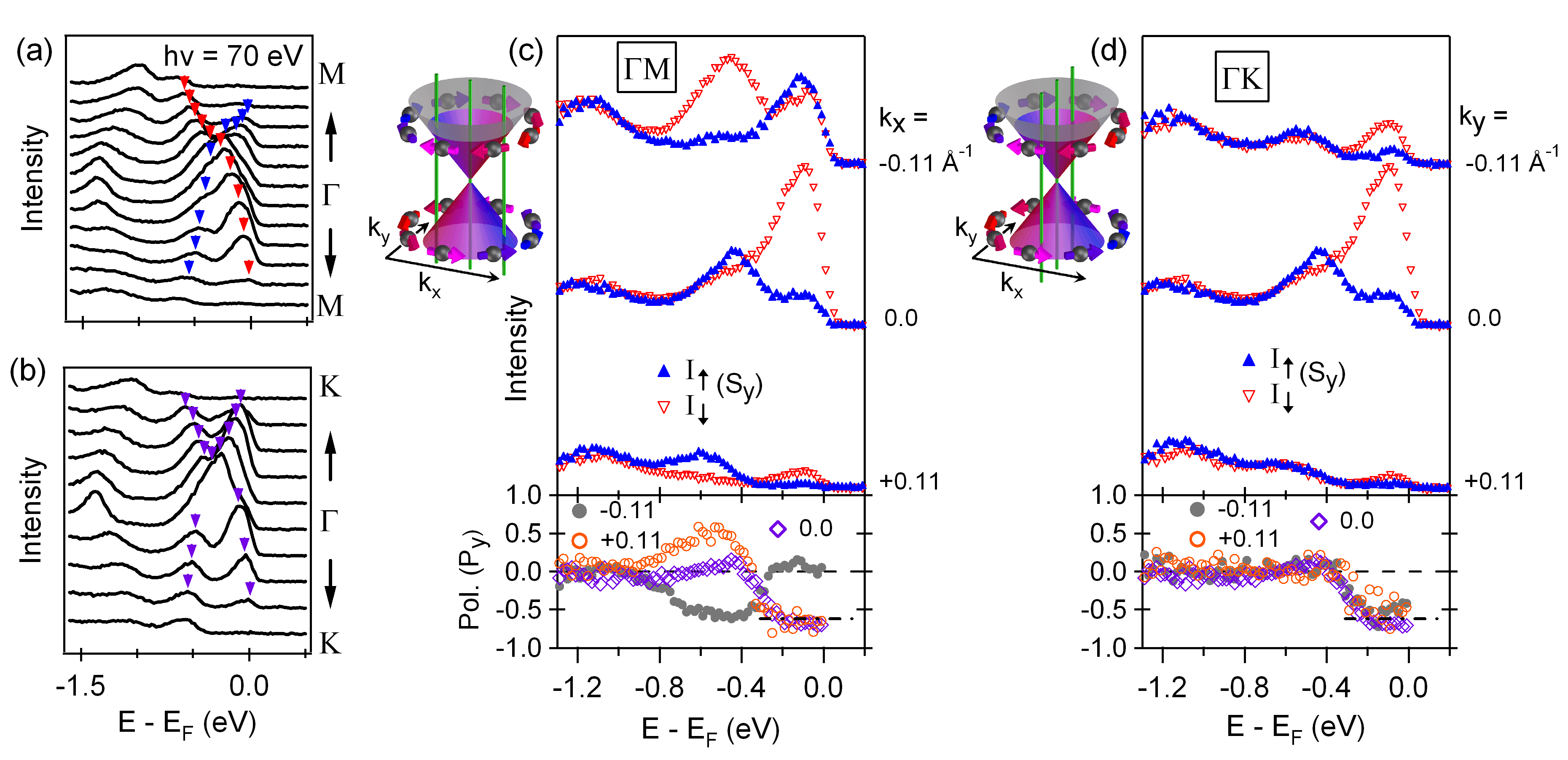}
\caption{\label{fig:hv70}(Color). (a) Spin-integrated EDCs as a function of momentum along $k_x$ ($\Gamma$M) with hv = 70 eV.  Peaks related to the TSS are marked by blue (spin-up) and red (spin-down) arrows according to their polarization along $\hat{y}$.  Other features are related to the bulk bands.
(b) Same as (a), except along $k_y$ ($\Gamma$K).  Peaks related to the TSS are marked by purple arrows, and are not expected to be polarized along $\hat{y}$.
(c) Spin-resolved EDCs along $k_x$ ($\Gamma$M).  Spin is measured along $\hat{y}$.  Approximate locations are shown by vertical green lines in left inset and the center and outer dots in the Fig.~\ref{fig:diagrams}(d) inset.  The corresponding spin polarization curves are shown in the lower panel.
(d) Same as (c), except along $k_y$ ($\Gamma$K).}
\end{figure*}

When considering only the photon polarization, $\boldsymbol{\hat{\epsilon}}$, photoelectron emission direction, $ \boldsymbol{\hat{k}}_e$, and the surface normal in the present geometry [see Fig.~\ref{fig:diagrams}(b)], $s$-polarized photons do not break the $yz$ mirror plane symmetry of the system as $p$-polarized photons do.
With two orthogonal mirror plans ($xz$ and $yz$), parity requires  $\boldsymbol{\vec P}_\mathrm{SME}$=0, as described above.

Note, however, the present solid-state surface introduces its own threefold symmetry which beaks the $yz$ mirror plane symmetry of the full system in the present geometry.
Thus, even in the case of $s$-polarized light, an SME-induced $P_y$ is not strictly forbidden by symmetry considerations.
Indeed, as proposed in Ref.~\onlinecite{Tamura1987} and observed on a Pt(111) surface \cite{Schmiedeskamp1988}, the resulting total symmetry allows for measurable SME-induced photoelectron polarization with normally incident $s$-polarized light.
Due to the role of the crystal structure in this particular variety of SME effect, the spin polarization is a strong function of sample azimuthal rotation with respect to the photon polarization \cite{Tamura1987,Schmiedeskamp1988}.
In an experiment using off-normal incidence of unpolarized light on Pt(111) and Au(111) surfaces \cite{Schmiedeskamp1991}, however, very little dependence of the photoelectron spin polarization on the sample azimuth was found.
It was argued that the SME-induced effects related to the off-normal incidence geometry are more general  than those related to the particular three-fold (111) crystal surface with normal photon incidence, as the latter require transitions into evanescent states within the bandgap \cite{Schmiedeskamp1988}.
A more thorough measurement of the photoelectron polarization with $s$-polarized light, as a function of sample azimuth (with respect to photon polarization), would be required to clarify this issue in the current case, but would be difficult due to the low photoemission cross section.

Just as the SMEs in the Bi $5d$ core levels are photon energy dependent, related SMEs in the near $E_F$ bands should be, as well.
Figure~\ref{fig:hv70} shows a dataset similar to Fig.~\ref{fig:SDb}, but taken with higher photon energy.
Panels (a) and (b) show spin-integrated EDC stacks along $k_x$ ($\Gamma$M) and $k_y$ ($\Gamma$K), respectively.
Again, the dispersing peaks of the TSS are marked by colored triangles according to their predicted spin polarization along $\mathbf{\hat{y}}$.
At this photon energy, the previous intense BVB peak at $E\sim -0.6$ eV is not present, possibly due to $k_z$ dispersion or suppression of the matrix elements, allowing clear observation of both the UDC and LDC dispersions.
Significant weight at $E_F$ from the BCB, however, is still present.

Panel (c) presents spin-resolved EDCs and $P_y$ curves analogous to Fig.~\ref{fig:SDb}(a).
Now the intrinsic spin texture of the LDC ($E\sim -0.5$ eV) is obvious even in the EDCs, with the sign of $P_y$ reversing through $\pm k_x$, as required by TRS.
The expected spin behavior of the UDC remains visible, but is now partly hidden by a strong negative polarization component at $E_F$ even at $k_x=0$.
As before, the $P_y(k_x=0)$ curve (purple diamonds) represents the $\mathbf{k}$-independent $\boldsymbol{\vec P}_\mathrm{SME}$ as a function of binding energy, which has changed with photon energy.
Here then, $\boldsymbol{\vec P}_\mathrm{SME} \approx 0$ at energies below the Dirac point, but quickly reaches $\sim -50\%$ at $-0.2 < E < E_F$, where the BCB is a strong source of intensity.
Similar to the $P_y$ curves of Fig.~\ref{fig:SDb}(a), the $P_y(k_x=\pm0.11$ \AA$^{-1})$ curves qualitatively appear reflected about the polarization at $k_x=0$.
A noticeable difference is that now $|P_y|$ near $E_F$ is not larger at $k_x=+0.11$\AA$^{-1}$ than at $k_x=0$.
This may be because the SME-induced polarized emission from the BCB contributes just as strongly to the total photoelectron polarization as the emission from the intrinsic polarized UDC quasiparticles.
This also explains the $P_y\approx 0$ measurement near E$_F$ at $k_x=-0.11$\AA$^{-1}$ as the two components would nearly cancel where the polarization component due to the intrinsic UDC reverses.

Panel (d) shows similar data along $k_y$.
As expected, due to the helical spin texture of the TSS (see inset), $P_y$ is independent of $k_y$; like Fig.~\ref{fig:SDb}(b), each $P_y$ curve shows the same form.
 
Thus, comparing Figs.~\ref{fig:SDb}(b)~and~\ref{fig:hv70}(d), we find that the $\mathbf{k}$-independent polarization features are strongly photon energy dependent, switching from $\sim$ +25\% in both bulk bands to $\sim$ -50\% in the BCB at photon energies of 36 and 70 eV, respectively.
This behavior is in support of an SME-induced origin.
Finally, note that this strong photon energy dependence is further strong evidence that the $\mathbf{k}$-independent polarization effects here cannot be instrumental artifacts.

\section{\label{sec:Sum}Summary}

To summarize, using a novel high-efficiency spin-resolved spectrometer to perform a thorough spin-ARPES study of Bi$_2$Se$_3$, we have observed two contrasting effects of spin-orbit coupling on the spin polarization of photoelectrons from a 3D topological insulator.
First, we observed spin polarization features with strong $\mathbf{k}$-dependence, consistent with TRS, that are due to the intrinsic TSS quasiparticle polarization and follow the expected spin-helical structure both above and below the Dirac point energy, even outside the bulk band gap.
Near $E_F$ we measured a very large polarization ($>$80\%) in the TSS.
Furthermore, we directly observed that this spin structure persists to room temperature.
Secondly, as a result of SMEs originating from strong spin-orbit coupling, a significant $\mathbf{k}$-independent polarization is observed, both in shallow core levels and near $E_F$ dispersive bands.
These observations clearly demonstrate the significant inequivalence of quasiparticle and photoelectron spin in these materials, strongly dependent on experimental parameters.
This suggests that full understanding is required before interpreting spin-ARPES data on these materials and extracting quantitative information regarding TSS quasiparticle polarization.
Finally, we note that in addition to the SME-induced polarization effects discussed above, there exist other effects (e.g. Refs.~ \onlinecite{Tamura1987,Henk2003,Kuemmeth2009}) that may result in differences between quasiparticle and photoelectron polarization.
Full relativistic one-step model matrix element calculations that include these polarization effects are thus required for realistic quantitative analysis of measured photoelectron polarization.
Indeed, advances have been made with tight integration of spin-ARPES results and such calculations (e.g. Refs.~\onlinecite{Henk2004,Barriga2009,Nuber2011}), and the current results stress the importance of a similar approach in the 3D topological insulators.

\begin{acknowledgments}

We thank J. E. Moore, O. V. Yazyev, A. Vishwanath, and H. Yao for helpful discussions.  We also thank G. Lebedev for work with the electron optics, J. Graf, C. G. Hwang, D. A. Siegel, S. D. Lounis, and W. Zhang for help with moving and installing the endstation, J. Sobota and J.J. Lee for experimental assistance, and A. Bostwick for help with software development.
This work was supported by the Director, Office of Science, Office of Basic Energy Sciences, Division of Materials Sciences and Engineering, of the U.S. Department of Energy under Contract No. DE-AC02-05CH11231 (Lawrence Berkeley National Laboratory) and Contract No. DE-AC02-76SF00515 (SLAC National Accelerator Laboratory).
The photoemission work was performed at the Advanced Light Source, Lawrence Berkeley National Laboratory, which is supported by the Director, Office of Science, Office of Basic Energy Sciences, of the U.S. Department of Energy under Contract No. DE-AC02-05CH11231.

\end{acknowledgments}

%bibliography{c:/Documents*and*Settings/chris/My*Documents/ChrisBiblio}
%%\bibliographystyle{Science}

\end{document}